\documentstyle[12pt]{article}

\font\tenrm=cmr10
\font\tenit=cmti10
\font\tenbf=cmbx10
\font\elevenbf=cmbx10 scaled\magstep 1
\font\elevenrm=cmr10 scaled\magstep 1
\font\elevenit=cmti10 scaled\magstep 1
\font\ninebf=cmbx9
\font\ninerm=cmr9
\font\nineit=cmti9

\begin{document}
\newcommand{\bibit}{\nineit}
\newcommand{\bibbf}{\ninebf}
\renewenvironment{thebibliography}[1]
{   \begin{list}{\arabic{enumi}.}
    {\usecounter{enumi} \setlength{\parsep}{0pt}
     \setlength{\itemsep}{2pt} \settowidth{\labelwidth}{#1.}
     \sloppy
    }}{\end{list}}

\parindent=3pc
\def\simlt{\mathrel{\lower2.5pt\vbox{\lineskip=0pt\baselineskip=0pt
           \hbox{$<$}\hbox{$\sim$}}}}
\def\simgt{\mathrel{\lower2.5pt\vbox{\lineskip=0pt\baselineskip=0pt
           \hbox{$>$}\hbox{$\sim$}}}}
\newcommand{\cm}{{\elevenit Commun. Math. Phys.}}
\newcommand{\pr}{{\elevenit Phys. Rev.}}
\newcommand{\prl}{{\elevenit Phys. Rev. Lett.}}
\newcommand{\pl}{{\elevenit Phys. Lett.}}
\newcommand{\np}{{\elevenit Nucl. Phys.}}

\begin{titlepage}
\begin{flushright}
\hfill{CPTH-C256.0693}\\[1mm]
June 1993
\end{flushright}
\vskip 1cm

\begin{center}{
{\tenbf REVIEW OF FREE-FERMIONIC 4D STRING MODELS\footnote{\ninerm Based
on talks presented at the SUSY-93 International Workshop on ``Supersymmetry
and Unification of Fundamental Interactions", Boston, March 29 - April 1,
1993, and at the XVI Kazimierz Meeting on Elementary Particle Physics,
Kazimierz, Poland, 24-28 May 1993\\}}
\vglue 1.0cm
{\tenrm I. ANTONIADIS\\}
\baselineskip=13pt
{\tenit Centre de Physique Th\'eorique, Ecole Polytechnique\\}
\baselineskip=12pt
{\tenit  91128 Palaiseau, France\\}
\vglue 0.8cm
{\tenrm ABSTRACT}}
\end{center}
\vglue 0.3cm
{\rightskip=3pc
\leftskip=3pc
\tenrm\baselineskip=12pt
\noindent
I review the main properties of four-dimensional strings constructed with
free-fermions on the world-sheet. In particular I discuss possible model
independent low energy predictions related to the existence of states
with fractional electric charges, the computation of the string
unification scale, the string model building, and the perturbative
approach to supersymmetry breaking which makes the spectacular prediction
of a new large dimension at the TeV scale.
\vglue 0.6cm}
\end{titlepage}
\baselineskip=14pt
\elevenrm
Despite the fact that progress in string theory was slow during the last
couple of years compared to the situation in the early days, it remains
actually the only known theory which contains quantum gravity, while
simultaneously it provides a framework of unification of all fundamental
interactions (gauge, scalar and Yukawa's).$^1$ At very short
distances, it modifies the structure of space-time, while at energies
lower than the Planck scale, all massive string excitations decouple
giving rise to an effective field theory which describes the dynamics of
massless modes. For the purpose of relevance to particle physics, there are
two main questions:
\hfil\\
(a) Is there a string vacuum which contains our low energy world as
described by the Standard model of strong and electroweak interactions?
\hfil\\
(b) Are there universal, model independent, string predictions at low
energies?

To answer these questions, one has first to compactify the ten-dimensional
heterotic superstring down to four dimensions. The string compactification
procedure consists of replacing the 6 left-moving and the 22 right-moving
internal coordinates by a (super) conformal field theory with central
charge $c=(9, 22)$. In the simplest constructions, the latter is a free
two-dimensional theory of bosons or fermions. Here, we restrict ourselves
to 4d free-fermionic strings which employ 18 left-moving and 44
right-moving 2d real fermions.$^{2,3}$ The various models then correspond
to the freedom of choosing the boundary conditions when the 2d fermions are
parallel transported around the string. These are constrained by multiloop
modular invariance which gives rise to well defined rules of construction.
The main advantage of these models is their simplicity since they are
obtained by tensoring Ising models which are solvable conformal field
theories, allowing a straightforward derivation of the spectrum and a
computation of the various couplings, correlation functions, etc. Their
disadvantage is that they describe particular points of the parameter space
(or their vicinity) which correspond to enhanced symmetric points of ${\bf
Z}_2 \times {\bf Z}_2$ orbifolds, and it is difficult to study generic
properties of continuous class of models.

The continuous parameters of four-dimensional strings consist of one
fundamental mass scale given by the Regge slope $\alpha'$ or
equivalently the Planck mass, $(\alpha')^{-1/2}\sim 10^{18}{\rm GeV}$, one
dimensionless coupling constant $g_4$ determined by the dilaton vacuum
expectation value (VEV), as well as several VEV's of scalar fields called
moduli which have zero potential. The relation between $\alpha'$, $g_4$
and the Newton's constant $G_N$ is:
\begin{equation}
G_N = \frac{g_4^2}{16\pi}\alpha'\ .
\label{scale}
\end{equation}
{}From the ten-dimensional point of view, the moduli VEV's correspond to
compactification parameters describing for instance the size and shape of
the internal manifold. An important property is that they are subject to
some exact discrete string symmetries called dualities which leave
invariant the effective low energy Lagrangian. The determination of all
scalar VEV's is a complicate dynamical problem, closely connected to the
mechanism of supersymmetry breaking, as well as to non perturbative
phenomena. In addition to the continuous parameters there is also a
certain number of discrete ones. Some of them will be discussed below.

Every consistent 4d string model contains in the massless spectrum the
graviton which appears together with a universal scalar field, the
dilaton, as well as a two-index antisymmetric tensor which is equivalent
to a pseudoscalar axion. Simultaneously, one obtains a gauge group which
can be as large as a group of rank 22.

Below, I discuss the general model independent properties and possible
predictions of 4d strings, which fall into 4 categories:
\hfil\\
1. ``Exotic" representations and/or fractional electric charges.
\hfil\\
2. Unification of couplings at a calculable unification scale.
\hfil\\
3. No unbroken continuous global symmetries implying, in particular, that
baryon and lepton number are in general violated and proton is unstable.
In fact, one of the serious potential phenomenological problems in string
model building is fast proton decay.
\hfil\\
4. A new large dimension at the TeV region is a spectacular prediction of
perturbative strings which relate its size with the scale of supersymmetry
breaking.
\vglue 0.6cm
{\elevenbf\noindent 1. Exotic representations and/or fractional
electric charges}
\vglue 0.4cm
Every gauge group factor is characterized by a positive integer $k$, the
level of the underlying 2d Kac-Moody algebra. This integer determines the
corresponding tree-level gauge coupling constant, $g=g_4 /{\sqrt k}$,
while simultaneously it restricts the possible massless matter
representations. For $k>1$, large ``exotic" representations are in general
present in the spectrum, like color octets, $SU(2)$ triplets, etc. On the
contrary, $k=1$ constructions, which include generic free-fermionic models,
guarantee that the only possible massless representations are fundamental
and antisymmetric of unitary groups or vectors and spinors of orthogonal
groups, which is consistent with the present experimental observations.
However, in this case, the observed electric charge quantization cannot be
imposed together with a value for the weak angle $\sin^2\theta_W =3/8$
at the unification scale.$^{4,5}$ Consequently, color singlet states with
fractional electric charges are unavoidable, unless the weak angle is
unacceptably small ($\sin^2\theta_W\le 3/20$). In free-fermionic models a
weaker charge quantization condition can always be imposed, namely that
the electric charges of color singlet states are half-integers.

The lightest fractionally charged particle is stable. Generically their
mass is in the TeV region, which seems to be experimentally excluded,
since an estimation of their relic abundancies contradicts the upper
experimental bounds by several orders of magnitude.$^6$ There are two
possible ways out. The first is to make them superheavy and use inflation
to suppress their abundancies as in the case of monopoles. The second and
most natural is to confine them by a gauge group of the hidden sector,
in the same way that QCD confines the fractionally charged quarks into
integrally charged bound states.$^7$ This possibility may lead to
interesting phenomenological consequences related to the existence of
long-lived ``crypto"-baryons.$^{8}$ It is also conceivable that the same
gauge group is also responsible for dynamical supersymmetry breaking via
gaugino and scalar condensation, in which case the confining scale must be
in the TeV region.$^9$
\vglue 0.6cm
{\elevenbf\noindent 2. Unification of couplings}
\vglue 0.4cm
Unlike field theories, 4d strings with $k=1$ imply an automatic
unification of all gauge couplings without the presence of a grand unified
group (GUT). Furthermore, the unification scale $M_{st}$ is calculable by
taking into account the string threshold corrections. In fact, at the
one-loop level one obtains:$^{10,11}$
\begin{equation}
\frac{16\pi^2}{g_i^2} = \frac{16\pi^2}{g_4^2} +
b_i\ln\frac{M_{st}^2}{\mu^2} + \Delta_i \ ,
\label{thr}
\end{equation}
where $\mu$ is an infrared cutoff, $b_i$ are the one-loop $\beta$-function
coefficients, and $\Delta_i$ define the finite threshold corrections given
by an integral over the complex modular parameter $\tau$ of the
world-sheet torus. It is important to realize that in string theory, as
well as in quantum field theory, the logarithmic infrared divergence,
associated with the running of low energy couplings, is due to the
integration over the massless particles. When the infrared divergence,
${\rm Im}\tau\rightarrow\infty$, is regularized and it is compared to the
field-theoretical $\overline{DR}$ scheme, one finds:$^{10}$
\begin{equation}
M_{st}\sim 5\times g_4 \times 10^{17} {\rm GeV}\ .
\label{mst}
\end{equation}

Since the unification scale is calculable, string theories, in contrast to
grand unified field theories, lead in general to two low energy
predictions. Both the weak angle and the strong coupling $\alpha_s$ are in
principle calculable in terms of the electromagnetic coupling constant and
the Planck mass. Unfortunately, assuming the particle content of the
minimal supersymmetric standard model below $M_{st}$, one obtains
$\sin^2\theta_W =.221$ and $\alpha_s =.203$ at $M_Z$, which are close but
clearly in contradiction with the experimental values. In fact $M_{st}$ in
Eq.~(\ref{mst}) is about a factor of 20 bigger than the desired value of
the unification scale which is consistent with the low energy couplings.
An obvious way to bridge this gap is by introducing an additional scale
which reduces the predictive power of the theory. One possibility is to
have some ``unnaturally" large modulus VEV, one to two orders of magnitude
larger than the Planck length, which gives rise to large threshold
corrections $\Delta_i$.$^{12}$ Alternatively, one may introduce additional
gauge non-singlet representations at some intermediate scale, which could
either be extra matter or enhance the gauge symmetry to a grand unified
type group.$^{13}$ At this stage, all proposed solutions are
not satisfactory, since our present understanding on the possible origin
of such intermediate scales is very limited.
\vglue 0.6cm
{\elevenbf\noindent 3. Model building}
\vglue 0.2cm
{\elevenit\noindent 3.1 General properties}
\vglue 0.1cm
As explained above, string theory does not lead to a unique way to
determine the vacuum, at least in perturbation theory. However, it
provides a framework for model building with rules which are much more
restrictive than in ordinary field theory. The simplest models correspond
to $k=1$ constructions and they are subject to several phenomenological
constraints. On the one hand they have to provide solutions to basic
outstanding problems, as the generation of the observed fermion mass
hierarchies, and on the other hand they should not suffer from obvious
diseases, as fast proton decay, or flavor changing neutral currents, etc.
These may be imposed using discrete symmetries at the level of the
Standard model, or proceeding through some grand unified group. In the
second case ordinary GUT's should be modified, since $k=1$ constructions
have no higgses in the adjoint or in higher self-conjugate representations
usually needed to break the GUT group.

In analogy with Calabi-Yau compactifications which lead to $E_6\times
E_8$ gauge symmetry, there is a class of free-fermionic 4d strings which
lead to an intermediate step of $SO(10)\times U(1)^n\times SO(16)$. The
gauge group is broken to a product of an ``observable" with a ``hidden"
sector (modulo the $U(1)$ factors), $G_{\rm obs}\times U(1)^n\times
G_{\rm hid}$, by 2d fermion number projections which correspond
to the Wilson line breaking of $E_6$ in Calabi-Yau models. These
projections also reduce the number of families to three. A general
property of these constructions is that every fermion generation forms a
spinor of $SO(10)$ instead of a {\bf 27} representation of $E_6$ and,
thus, it contains just one more state, the right-handed neutrino, besides
the known Standard model content. There are three distinct possibilities
for $G_{\rm obs}$, associated to the three different subgroups of $SO(10)$
which can be broken to the Standard model using higgses in low dimensional
representations: the flipped $SU(5)\times U(1)$,$^{14,7}$ the Pati-Salam
left-right symmetric model $SU(4)\times SU(2)_L\times SU(2)_R$,$^{15}$ or
directly the Standard-like model $SU(3)\times SU(2)_L\times U(1)_Y\times
U(1)'$.$^{16}$

The general strategy of this procedure, before introducing the breaking
of space-time supersymmetry, is focusing two main issues: The gauge
symmetry breaking and the fermion mass problem. The gauge symmetry
breaking contains four steps:
\hfil\\
(i) The breaking of the additional $U(1)$ factors. This is achieved
because one particular linear combination $U(1)_A$ appears often to be
anomalous at the tree-level. The anomaly is canceled by a generalization
of the Green-Schwarz mechanism in four dimensions, implying that the axion
is absorbed by the $U(1)_A$ gauge field which becomes massive at the
one-loop level.$^{17}$ Furthermore, the corresponding $U(1)_A$ $D$-term is
modified by an additional ``constant" proportional to the anomaly ${\rm
Tr}Q_A$:
\begin{equation}
D_A = \sum_i Q_A^i|\Phi_i|^2 + \frac{{\rm Tr}Q_A}{96\pi^2\alpha'}\ ,
\label{dterm}
\end{equation}
where $\Phi_i$ are the various scalars. As a result, some of the
previously flat directions are destroyed, while the supersymmetric
minimization conditions of $F$ and $D$-flatness are satisfied by non
vanishing VEV's of scalar fields which in general break all extra $U(1)$'s
at a calculable scale $M_A$. For typical values of ${\rm Tr}Q_A\sim
1-100$, one finds a scale which is two to one orders of magnitude less
than the string scale, $M_A\sim 10^{16}-10^{17}$ GeV.
\hfil\\
(ii) The breaking of the GUT-type group $G_{\rm obs}$ to the Standard model
by a Higgs mechanism, at a scale $M_{\rm GUT}\sim 10^{16}$ GeV which
could have the same or a different origin than $M_A$.
\hfil\\
(iii) The condensation of the hidden group $G_{\rm hid}$ at some
intermediate scale $\Lambda_{\rm cond}$, which should also confine the
fractionally charged states and guarantee the observed electric charge
quantization. The hidden group may also be related with the dynamical
breaking of space-time supersymmetry via gaugino condensation.$^{18,9}$
\hfil\\
(iv) The usual $SU(2)\times U(1)$ breaking which needs one pair of massless
Higgs doublets. This is in general a problem in any construction, since
vector-like representations tend to become massive.

Concerning the fermion masses, the strategy is the following:$^7$
\hfil\\
Only one generation has trilinear Yukawa couplings and aquires a mass at
the lowest order in the $\alpha'$-expansion. Because of the underlying
$SO(10)$ symmetry one obtains $m_b=m_{\tau}$ at the unification scale,
which leads to a successful prediction for the bottom to tau mass ratio at
low energy, after taking into account the renormalization group evolution.
Furthermore, the top Yukawa coupling $\lambda_t$ is proportional to the
gauge coupling at $M_{st}$ implying that the top is in general heavy. In
fact an upper bound can be obtained, corresponding to the limiting case
$\lambda_t =\lambda_b =\lambda_{\tau}={\sqrt 2}g_4$ at $M_{st}$ which is
valid in the absence of any mixing:$^7$
\begin{equation}
m_t\simlt 180 {\rm GeV}\ .
\label{top}
\end{equation}
All other masses and mixings involving the other two generations appear
in $\alpha'$-corrections which correspond to non renormalizable terms in
the effective superpotential. Consequently, the corresponding Yukawa
couplings are naturally suppressed by powers of $M_A(\alpha')^{1/2}$, or
$M_{\rm GUT}(\alpha')^{1/2}$, generated by the various scalar VEV's.
\vglue 0.2cm
{\elevenit\noindent 3.2 $SU(5)$ unification}
\vglue 0.1cm
As explained above, in the context of fermionic constructions described
here, ordinary $SU(5)$ unification is not possible. Its simplest variation
is flipped $SU(5)$, where the up and down antiquark triplets are
interchanged between the ${\bf 10}$ and ${\bf{\bar 5}}$ representations,
while simultaneously the right handed electron and neutrino are
interchanged between the ${\bf 10}$ and a singlet. Thus, every generation
forms an $SO(10)$ spinor. Note that the electric charge is not anymore an
$SU(5)$ generator but it mixes with an additional $U(1)$ factor which must
be present.$^{19}$ The gauge group $SU(5)\times U(1)$ is broken to the
Standard model by the VEV of a ${\bf 10}+{\bf\overline{10}}$
representation along the neutral $\nu^c,{\bar\nu}^c$ direction.
Furthermore, one needs a Higgs in the ${\bf 5}+{\bf{\bar 5}}$
representation, embedded in a vector of $SO(10)$, to break the electroweak
symmetry.

One serious problem of $SU(5)$ unification, which is a manifestation of
gauge hierarchy, is that the Higgs triplets lying inside ${\bf
5}+{\bf{\bar 5}}$ must be superheavy because they mediate fast proton
decay, while the remaining Higgs doublets must be light since they play
the role of Standard model higgses. In the minimal $SU(5)$ this
triplet-doublet splitting requires a severe fine tuning, while flipped
$SU(5)$ offers a natural solution to the problem.$^{14}$ In fact, due to
an allowed superpotential coupling between the two set of higgses, ${\bf
10}-{\bf 10}-{\bf 5}$, the anti-triplet of ${\bf 10}$ is paired with the
triplet of ${\bf 5}$ and they become superheavy when $SU(5)$ is broken,
while the Higgs doublets remain massless. A further consequence of this
mechanism is the absence of dimension 5 operators for proton decay,
implying a life-time of the order of $10^{36}$ years with dominant decay
modes $p\rightarrow{\bar\nu}\pi^+$, or $e^+\pi^0$.

A minimal $SU(5)\times U(1)$ field-theoretical toy model should contain,
in addition to the three families of $SO(10)$ spinors decomposed as
$F_i=({\bf 10},1/2)$, ${\bar f}_i=({\bf{\bar 5}},-3/2)$, and $l^c_i=({\bf
1},5/2)$, $i=1,2,3$, a pair of GUT higgses $H=({\bf 10},1/2)$ and ${\bar
H}=({\bf\overline{10}},-1/2)$, a pair of light higgses $h=({\bf 5},-1)$
and ${\bar h}=({\bf{\bar 5}},1)$ forming a vector of $SO(10)$, as well as
three $SO(10)$ singlets $\phi_i=({\bf 1},0)$. The most general trilinear
superpotential, which is invariant under the discrete ${\bf Z}_2$ symmetry
$H\rightarrow -H$, is:
\begin{equation}
W = \lambda_d FFh + \lambda_u F{\bar f}{\bar h} + \lambda_e{\bar f}l^ch
+ \lambda_{\nu} F{\bar H}\phi + \lambda_{\phi}\phi^3
+ \lambda_{\mu}h{\bar h}\phi
+ \lambda_1 HHh + \lambda_2{\bar H}{\bar H}{\bar h}\ ,
\label{w}
\end{equation}
where the generation indices were omitted for simplicity. The first five
couplings in Eq.~(\ref{w}) give rise to the fermion masses, while the
last two are responsible for the triplet-doublet splitting. In particular,
$\lambda_{d,u,e}$ generate masses for down quarks, up quarks and charged
leptons, respectively, while $\lambda_u$ gives also Dirac masses to
neutrinos. In contrast to the ordinary $SU(5)$, down quarks and leptons
belong to different multiplets and get masses from different Yukawa
couplings. As a result, one looses the relation $m_d=m_e$ at $M_{\rm
GUT}$, which is successful for the third generation but problematic for
the other two. Instead, one now obtains $m_{\nu}=m_u$. However, the
couplings $\lambda_{\nu}$ and $\lambda_{\phi}$ generate a generalized
see-saw mechanism for the right-handed neutrinos which form superheavy
Dirac eigenstates with the singlets $\phi$. Diagonalizing the resulting
mass matrix of $\nu_i$, $\nu^c_i$ and $\phi_i$, one also finds tiny
Majorana masses for the left-handed neutrinos of the order of
$m_u^2<\phi>/M_{\rm GUT}^2$.

Although simple and economic, the field theoretical $SU(5)\times U(1)$
model is not a real grand unified theory. It introduces two independent
gauge coupling constants, and one looses the mass relation $m_b=m_{\tau}$.
On the contrary, in the context of 4d strings gauge coupling unification
is automatic, while the successful relation $m_b=m_{\tau}$ is retained as
a consequence of the underlying $SO(10)$ structure in the trilinear
couplings of these constructions. This symmetry is broken in the massive
string sector so that the corresponding physically unrealistic relations
for the lighter generations, which acquire masses through
non-renormalizable interactions, are expected to be violated.

An explicit example of a three-generation flipped $SU(5)$ model with the
above properties was derived using the fermionic constructions of 4d
strings.$^7$ The total gauge group is $G_{\rm obs}\times U(1)^4\times
G_{\rm hid}$, with $G_{\rm obs}=SU(5)\times U(1)$ and $G_{\rm hid}
=SO(10)\times SU(4)$. The massless spectrum consists of 4 complete
families + 1 antifamily, 4 pairs of ${\bf 5}+{\bf{\bar 5}}$ higgses, 10
singlets charged under $U(1)^4$, 5 neutral singlets, as well as ``hidden"
matter which contains 5 $SU(5)\times U(1)$ singlets transforming as
$\{({\bf 10},{\bf 1}) + ({\bf 1},{\bf 6})\}$ under $G_{\rm hid}$, and 6
$SU(5)$ singlets transforming as $\{({\bf 1},{\bf 4}) + ({\bf 1},{\bf{\bar
4}})\}$ and having fractional electric charges $\pm 1/2$. The pair of GUT
higgses $H+{\bar H}$ is provided by one of the families and the
antifamily, while one linear combination of the four $U(1)$'s appears to
be anomalous. Also, the trilinear superpotential is easily derived and it
exhibits the desired properties. As a result, the analysis of gauge
symmetry breaking follows the general description presented above.

The non-renormalizable terms in the effective superpotential were
computed and analyzed up to 8th order in the $\alpha'$-expansion by two
different groups.$^{20,21}$ It turns out that there is an appropriate
choice of scalar VEV's which leads to reasonable phenomenology. In
particular, all triplets become superheavy, while there remain one or two
pairs of massless doublets. Furthermore, one obtains fermion masses having
the correct orders of magnitude. Despite many interesting features, this
model has two main defects. The generic problem of the unification scale
which cannot be lowered easily using the field content present in the
spectrum, and the problem of fixing the arbitrariness in the choice of
scalar VEV's used to explain the fermion mass hierarchies.
\vglue 0.2cm
{\elevenit\noindent 3.3 Pati-Salam unification}
\vglue 0.1cm
A different maximal subgroup of $SO(10)$, which can be broken to the
Standard model without adjoint higgses, is $SU(4)\times SU(2)_L\times
SU(2)_R$. The fermion generations form again spinors of $SO(10)$, which
are now decomposed as $F_i=({\bf 4},{\bf 2},{\bf 1})$ and ${\bar F}_i=
({\bf{\bar 4}},{\bf 2},{\bf 1})$, containing the left-handed and
right-handed components, respectively. The gauge symmetry is broken to the
Standard model by a pair of higgses $H=({\bf 4},{\bf 1},{\bf 2})$ and
${\bar H}=({\bf{\bar 4}},{\bf 1},{\bf 2})$. Furthermore, the two
electroweak higgses form the representation $h=({\bf 1},{\bf 2},{\bf
2})$, which emerges from the vector of $SO(10)$ together with a sextet
$D_6=({\bf 6},{\bf 1},{\bf 1})$.

The properties of this model are very similar to those of flipped
$SU(5)$. On the one hand, the couplings $\lambda_{1,2}$ in the
superpotential of Eq.~(\ref{w}), realizing the triplet-doublet splitting,
are replaced by $\lambda_1 HHD_6 + \lambda_2 {\bar H}{\bar H}D_6$ which
pairs the Higgs triplets with those of sextets into superheavy states. On
the other hand, the first three couplings $\lambda_{d,u,e}$ are replaced
by a single term $\lambda F{\bar F}h$ which gives masses to all fermions.
Thus, one recovers the $SO(10)$ relation $m_d=m_e$ together with
$m_u=m_{\nu}$. Finally, the see-saw mechanism for the right-handed
neutrinos works in the same way as before through the couplings
$\lambda_{\nu ,\phi}$. A first instance of a string derived model with
these properties was presented in Ref.15.
\vglue 0.6cm
{\elevenbf\noindent 4. Perturbative supersymmetry breaking}
\vglue 0.4cm
One of the most important problems in string theory, which establishes the
connection with our low energy world, is the breaking of space-time
supersymmetry. In the context of the effective field theory, this
breaking can be parameterized by introducing additional low energy
parameters which determine for instance the mass-splittings inside
supermultiplets. On the contrary, in string theory, there is no
independent parameter related to the scale of supersymmetry breaking and
the situation is very restrictive. It turns out that in perturbation
theory this scale is necessarily linked with the size of an internal
compactified dimension.$^{22}$ Thus, since supersymmetry breaking must
occur at energies close to the weak scale to protect the gauge hierarchy,
4d strings predict a new dimension in the TeV region!

Such a large dimension appears to have a lot of theoretical problems. The
most serious is that all couplings blow up very rapidly above the
decompactification scale, just by naive dimensional analysis since the
effective field theory becomes higher dimensional, and perturbation
theory breaks down. From the four-dimensional point of view the problem
appears with the production of an infinite tower of Kaluza-Klein (KK)
excitations, corresponding to the components of the momentum along the
compactified directions, which increase indefinitely the
$\beta$-functions.

This is the main reason why this possibility had not received attention
and supersymmetry breaking is usually assumed to occur
non-perturbatively. However, it was recently suggested that there is a
way out of the large coupling problem in a particular class of 4d string
models, which include orbifold compactifications.$^{23}$ This opens the
exciting possibility of lowering part of the massive string spectrum at
energies accessible to future accelerators. The main observation is that
in these models the KK-excitations are organized in multiplets of $N=4$
(spontaneously broken) supersymmetry, leading to cancellations of large
corrections among particles of different spin. In particular
$\beta$-functions remain small and gauge couplings evolve logarithmically
up to the Planck scale, despite the presence of an infinite tower of
modes which fill up the ``desert". Furthermore, the chiral character of
the theory implies the existence of a new class of states, called twisted,
which have no field theory analog as they are not accompanied by
KK-excitations.

The simplest realization of the minimal supersymmetric standard model in
the context of this mechanism was worked out recently and it was shown to
be very restrictive, even in the absence of an explicit string
construction.$^{24}$ Quarks and leptons should be identified with twisted
states having no KK-excitations. Moreover, at the tree-level, the only
source of supersymmetry breaking is a common gaugino mass, $m_{1/2}=1/2R$,
determined by the compactification radius $R$, while all scalar masses and
trilinear scalar couplings vanish, $m_0=A=0$. On the other hand, there is
a simple choice for the Higgs sector, which offers a natural solution to
the $\mu$-problem. The second Higgs doublet, characteristic of the
supersymmetric standard model, can be identified with the first massive
KK-mode of the first Higgs doublet carrying opposite hypercharge. In this
way, one obtains a Higgs mixing superpotential term with a coupling $\mu$
equal to the gaugino mass, together with a non-universal soft breaking
Higgs potential:
\begin{equation}
m_{1/2}=\mu=\frac{1}{2R}\ \ \ \ \ m_0=A=B=0\ \ \ \ \
V_{\rm soft}=-\mu^2 h_1^2 + 3\mu^2 h_2^2\ .
\label{sb}
\end{equation}
Note that although supersymmetry is broken, its breaking scale determined
by $R$ is arbitrary. In fact $R$ is given by the VEV of a modulus field
which corresponds to a flat direction of the tree-level scalar potential,
even in the presence of supersymmetry breaking. This situation is
reminiscent of the no-scale supergravity models.$^{25}$

This mechanism of supersymmetry breaking has an additional interesting
feature which allows the dynamical determination of scales by minimizing
the full one-loop effective potential with respect to the modulus and
Higgs fields. It is the absence of quadratic divergences in the
one-loop cosmological constant ($Str{\cal M}^2=0$) due to the underlying
$N=4$ supersymmetry of the massive modes. It follows that the effective
potential of the Higgs and modulus fields is independent of the Planck
mass at the renormalizable level. As a result, a new scale $Q_0$ is
dynamically generated through the running of the renormalization group
equations, defined as the energy where the mass-squared of the Higgs
becomes negative and triggers electroweak symmetry breaking. Both the
Higgs VEV and $R^{-1}$ are then proportional to $Q_0$ which can be
hierarchically smaller than the string scale, depending on the value of
the top Yukawa coupling.

Starting with the initial conditions of Eq.~(\ref{sb}) and replacing the
minimization with respect to the modulus by the phenomenological condition
on the value of the $Z$-mass, one is left with one free parameter, the top
Yukawa coupling. An analysis of both theoretical and experimental
constraints shows that there is an allowed region,
\begin{equation}
140\ {\rm GeV}\simlt m_t\simlt 155\ {\rm GeV}\ ,
\label{range}
\end{equation}
where the whole particle spectrum is given as a function of the top
mass.$^{24}$ The lower bound comes from the present experimental limits on
supersymmetric Higgs detection, while at the upper bound the value of the
supersymmetry breaking scale goes to infinity. In the lower bound the
masses of scalar quarks and gluinos are lying in the range of 200-300 GeV,
while the masses of scalar leptons, neutralinos, charginos and higgses in
the range of 50-150 GeV. Also, the compactification scale is around 200
GeV. Finally, the values of $\tan\beta$ are in the range between 1.7 and
4, while the lightest supersymmetric particle is a neutralino in the whole
region of Eq.~(\ref{range}).

A characteristic signature of this mechanism is the existence of
Kaluza-Klein excitations for gauge bosons and higgses at low energy. In
particular the lightest one is an excited photon with mass $1/R$, and it
could be accessible to future accelerators with a very clear signal in
the $l^+l^-$ channel. The present experimental limits on the size of
extra dimensions follow from the analysis of effective four-fermion
interactions induced by the exchange of KK-modes, and imply that $R^{-1}$
can be as low as 200 GeV.$^{24}$
\vglue 0.6cm
{\elevenbf \noindent Acknowledgements \hfil}
\vglue 0.4cm
This work was supported in part by a CNRS-NSF collaborative grant, and in
part by the EEC contracts SC1-0394-C, SC1-915053 and SC1-CT92-0792.
\vglue 0.6cm
{\elevenbf \noindent References \hfil}
\vglue 0.4cm

\end{document}